\documentclass[11pt]{article}
\usepackage{graphicx,amsmath,amsthm,mathrsfs,amssymb}
\usepackage[usenames]{color}
\usepackage{ulem,cite}
\usepackage{authblk}

\newcommand{\ee}{\end{equation}}

\newcommand{\reff}[1]{(\ref{#1})}
\newcommand{\beq}{\begin{equation}}
\newcommand{\eeq}[1]{\label{#1}\end{equation}}
\newcommand{\beqa}{\begin{eqnarray}}
\newcommand{\eea}{\end{eqnarray}}
\newcommand{\eeqa}[1]{\label{#1}\end{eqnarray}}
\newcommand{\beg}{\begin{equation*}}
\newcommand{\eeg}{\end{equation*}}

\newcommand{\bsplit}{\begin{split}}
\newcommand{\esplit}{\end{split}}

\title{Palatini formulation of pure $R^2$ gravity yields Einstein gravity with no massless scalar}

\author[1]{Ariel Edery\thanks{aedery@ubishops.ca}}
\author[2]{Yu Nakayama\thanks{yu.nakayama@rikkyo.ac.jp}}

\affil[1]{Department of Physics, Bishop's University, \\2600 College Street, Sherbrooke, Qu\'{e}bec, Canada, J1M 1Z7 .\vspace{8mm}}
\affil[2]{Department of Physics, Rikkyo University, Toshima, Tokyo 171-8501, Japan}
\date{}
\begin{document}
\maketitle
\begin{abstract}
Pure $R^2$ gravity has been shown to be equivalent to Einstein gravity with non-zero cosmological constant and a massless scalar field. We show that the Palatini formulation of pure $R^2$ gravity is equivalent to Einstein gravity with non-zero cosmological constant as before but with no massless scalar field. This is an important new development because the massless scalar field is not readily identifiable with any known particle in nature or unknown particles like cold dark matter which are expected to be massive. We then include a non-minimally coupled Higgs field as well as fermions to discuss how the rest of the standard model fields fit into this paradigm. With Higgs field, Weyl invariance is maintained by using a hybrid formalism that includes both the Palatini curvature scalar $\mathcal{R}$ and the usual Ricci scalar $R$.
\end{abstract}
\section{Introduction}
Pure $R^2$ gravity (i.e. $R^2$ gravity with no extra $R$ term) possesses a symmetry that is larger than scale symmetry and smaller than full Weyl symmetry. This was dubbed restricted Weyl symmetry \cite{Edery:2014nha} as the action is invariant under $g_{\mu\nu} \to \Omega^2(x) g_{\mu\nu}$ with the condition that $\Box \Omega=0$. This theory was shown to be equivalent to Einstein gravity with non-zero cosmological constant and a  massless scalar field \cite{Lust1,Lust2,Edery:2015wha,Lust3,YN3} (see also \cite{Ghilencea} for a Weyl geometry approach.)  The massless scalar is identified as the Nambu-Goldstone boson of the spontaneously broken restricted Weyl symmetry \cite{YN3}.  

The Palatini formulation \cite{Arnowitt:1962hi}\footnote{For use of the Palatini formalism in various contexts see references \cite{Bauer}-\cite{Karam}.} of pure $R^2$ gravity is Weyl invariant \cite{Borowiec:1996kg} and we show in this paper that it is equivalent to Einstein gravity with non-zero cosmological constant as before but no massless scalar field appears in contrast to the metric formulation. This is an important and positive development; the presence of a massless scalar is not an issue theoretically but as far as we know, there is no evidence for such a particle in nature. In particular, it cannot act as a cold dark matter candidate which is expected to be massive. We then introduce into the action a non-minimally coupled massless Higgs field. To maintain Weyl invariance, we use a hybrid formalism where the action includes both the Palatini curvature scalar $\mathcal{R}$ and the usual Ricci scalar $R$. 
We show that this action is equivalent to Einstein gravity with cosmological constant and a non-minimally coupled \textit{massive} Higgs field. Again, no massless scalar field appears. The original massless Higgs becomes massive and can now be taken to be the doublet of the standard model. The rest of the standard model fields can be readily included into this paradigm. In particular fermions can be included via the torsion-free Levi-Civita spin connection $\omega_{\mu}^{ab}$ constructed from the vielbein. A separate Palatini spin connection $\Omega_{\mu}^{ab}$ is also introduced and applied to the gravity sector. The main difference between the two is that the Levi-Civita spin connection changes under a Weyl transformation while the Palatini spin connection remains unchanged. The use of these two connections in the action is prescribed by the principle that the action is Weyl invariant.    

\section{Palatini formulation of pure $R^2$ gravity}

We show that the Palatini formulation of pure $R^2$ gravity, upon gauging the Weyl symmetry, is equivalent to Einstein gravity with non-zero cosmological constant. First note that pure $R^2$ gravity in the Palatini formulation in four dimensions is Weyl invariant \cite{Borowiec:1996kg} unlike in the metric formulation, which has only restricted Weyl symmetry \cite{Edery:2014nha,Edery:2015wha}. The Palatini action is given by
\begin{align}
S = \int d^4x \sqrt{-g} \alpha \mathcal{R}^2 \ . 
\end{align}
Here, the Palatini scalar curvature $ \mathcal{R}$ is given by 
\begin{align}
 \mathcal{R} = g^{\mu \sigma} (\partial_\nu \Gamma^{\nu}_{\mu\sigma} - \partial_\sigma \Gamma^{\nu}_{\mu\nu} + \Gamma^{\nu}_{\alpha \nu} \Gamma^{\alpha}_{\mu\sigma} - \Gamma^{\nu}_{\alpha\sigma}\Gamma^{\alpha}_{\mu\nu})
\end{align}
where we treat $g_{\mu\nu}$ and $\Gamma^{\sigma}_{\mu\nu}$ as independent variables (i.e. $\Gamma^{\sigma}_{\mu\nu}$ is not the Christoffel connection at this point). The above action is invariant under the Weyl transformation 
\begin{align}
g_{\mu\nu} &\to \Omega^2(x) g_{\mu\nu} \cr
\Gamma^{\mu}_{\nu\rho} &\to \Gamma^{\mu}_{\nu\rho} \ .
\end{align}
Note that in contrast to the metric formulation, the connection is chosen to be unchanged under a Weyl transformation. Thus the scalar curvature transforms as
\begin{align}
\mathcal{R} \to \Omega^{-2}(x)\mathcal{R}\ 
\end{align}
while the metric Ricci scalar, denoted by $R$, transforms as 
\begin{align}
{R} \to \Omega^{-2} R - 6\Omega^{-3} \Box \Omega \ .
\end{align}
It is now obvious that the pure $\mathcal{R}^2$ action is Weyl invariant in the Palatini formulation because we use $\mathcal{R}$ instead of $R$.

Let us rewrite the Palatini $R^2$ action by introducing an auxiliary field $\varphi(x)$ as
\begin{align}
S_1 &= \int d^4x \sqrt{-g}\left( -\alpha(c_1\varphi + \mathcal{R})^2 + \alpha \mathcal{R}^2 \right) \cr
    &= \int d^4x \sqrt{-g}\left(-c_1^2\alpha\varphi^2 -2\alpha c_1 \varphi \mathcal{R} \right) \ , \label{aux}
\end{align}
where $c_1$ is an arbitrary (non-zero) number. The above action is still Weyl invariant if $\varphi$ transforms as $\varphi \to \varphi/\Omega^2$ when $g_{\mu\nu} \to \Omega^2(x) g_{\mu\nu}$.

We now perform the Weyl transformation
\begin{align}
g_{\mu\nu} &\to \varphi^{-1} g_{\mu\nu} \cr
\sqrt{-g} &\to \varphi^{-2} \sqrt{-g} \cr
\mathcal{R} &\to \varphi \mathcal{R} \ .
\label{WT}
\end{align}
The resulting action is
\begin{align}
S_E &= \int d^4x \sqrt{-g} \left(-c_1^2\alpha - 2\alpha c_1 \mathcal{R} \right) \ . \label{PE}
\end{align}
This is nothing but Einstein gravity with non-zero cosmological constant in the Palatini formulation. Varying the action with respect to $\Gamma^{\mu}_{\rho\sigma}$ gives the metric compatibility condition 
\begin{align}
D_\rho g^{\mu\nu} = 0 \ ,
\end{align}
which now identifies $\Gamma^{\mu}_{\rho\sigma}$ with the Christoffel  connection (with the assumption of the symmetric connection)
\begin{align}
\Gamma^{\mu}_{\rho\sigma} = \frac{1}{2}g^{\mu\nu} (\partial_\rho g_{\sigma \nu} + \partial_{\sigma} g_{\rho \nu} - \partial_{\nu} g_{\rho\sigma}) \ .
\end{align}
 Varying the action with respect to $g_{\mu\nu}$ then gives the Einstein equations with cosmological constant. The Einstein action \reff{PE} is not Weyl invariant as the symmetry is now spontaneously broken. This occurs because the Weyl transformation \reff{WT} is not valid (i.e. is singular) for $\varphi=0$ and hence excludes the solution $\mathcal{R}=0$. With non-zero $\mathcal{R}$, the Weyl invariance is spontaneously broken.  

The point here is that the auxiliary scalar $\varphi$ disappears completely due to the Weyl invariance.\footnote{When we break the Weyl invariance e.g. by adding the linear $\mathcal{R}$ term in the action, $\varphi$ does not disappear \cite{Antoniadis:2018ywb}.} We should regard $\varphi$ as a would-be Nambu-Goldstone boson for the spontaneous breaking of the Weyl symmetry \cite{YN3}. However, because of the Weyl invariance, it does not appear in the final action. At this point, it is better to gauge the Weyl symmetry, so that we get rid of $\varphi$. There is no Nambu-Goldstone boson for spontaneously broken gauge symmetry.
Thus, the Weyl gauged $R^2$ gravity in the Palatini formulation is completely equivalent to the Einstein gravity with (arbitrary but non-zero) cosmological constant once we gauge the Weyl symmetry. 

An alternative way to look at the fate of $\varphi$ is that $\varphi$ transforms under the Weyl symmetry as
\begin{align}
\varphi \to \Omega^{-2} \varphi \ .
\end{align}
Thus when we gauge the Weyl symmetry, one can set $\varphi$ as we like to fix the gauge. Here we take $\varphi = 1$ in \eqref{aux} to obtain \eqref{PE}.

\section{Inclusion of Higgs field and fermions}
We may generalize the construction by introducing the Higgs field as we did in our previous papers \cite{Edery:2015wha,YN3}. Given the mechanism of decoupling of $\varphi$ in pure $R^2$ case, the crucial assumption here is we impose the Weyl invariance and gauge it. The (most) general Weyl invariant action we start with is
\begin{align}
S_1 = \int d^4x\sqrt{-g} \left(\alpha \mathcal{R}^2 - \xi \mathcal{R}|\Phi|^2 -\frac{1}{6} R |\Phi|^2 - (\partial_\mu \bar{\Phi} \partial^\mu \Phi) - \lambda|\Phi|^4 \right) \ . \label{Higgs}
\end{align}
We stress that $\mathcal{R}$ is the Palatini scalar curvature while $R$ is the Ricci scalar constructed out of the metric tensor (and its derivatives) and they are different (at this point). 
The non-minimal coupling of the Higgs field to $R$ is fixed (i.e. $1/6$)  by the Weyl invariance while the non-minimal coupling constant $\xi$, which couples to the Palatini curvature $\mathcal{R}$, is arbitrary. 

At this point, we should stress that we depart slightly from the original philosophy of Palatini that avoids the use of the (Riemann) curvature in the action. Our philosophy rather is to impose the Weyl invariance, and with the Palatini curvature, \eqref{Higgs} is the most general Weyl invariant unitary action that we can construct.\footnote{The Weyl invariance allows other terms such as Weyl squared, but generally it predicts ghost modes. See e.g. \cite{Alvarez:2018nxc}\cite{BeltranJimenez:2019acz} for a recent study in the Palatini formulations.} In particular, it is impossible to introduce the Weyl invariant kinetic term for scalar fields with only the Palatini curvature.

We introduce the auxiliary field $\varphi$ to rewrite the action as 
\begin{align}
S_1 = \int d^4x\sqrt{-g} \left( -\alpha\left( c_1\varphi + \mathcal{R} + \frac{c_2}{\alpha}|\Phi|^2 \right)^2 + \alpha \mathcal{R}^2 - \xi \mathcal{R}|\Phi|^2  - \frac{1}{6}{R}|\Phi|^2  - (\partial_\mu \bar{\Phi} \partial^\mu \Phi) - \lambda|\Phi|^4 \right) \  
\end{align}
with arbitrary constants $c_1$ and $c_2$.
After the Weyl transformation, 
\begin{align}
g_{\mu\nu} &\to \varphi^{-1} g_{\mu\nu} \cr
\sqrt{-g} &\to \varphi^{-2} \sqrt{-g} \cr
\mathcal{R} &\to \varphi \mathcal{R} \cr
\Phi &\to \varphi^{1/2} \Phi\cr
R & \to \varphi R - 6\varphi^{3/2} \Box \varphi^{-1/2} 
\end{align}
the action becomes
\begin{align}
S = \int d^4x \sqrt{-g} \Big(-\alpha c_1^2 -2\alpha c_1 \mathcal{R} - \partial_\mu \bar{\Phi} \partial^\mu \Phi &- 2c_1c_2 |\Phi|^2 \nonumber\\ &- (\alpha^{-1}c_2^2+\lambda)|\Phi|^4 - (\xi +2c_2) \mathcal{R} |\Phi|^2 -\frac{1}{6} {R} |\Phi|^2 \Big)\,.
\end{align}
Note that $\varphi$ does not appear in the final action thanks to the Weyl invariance of the original action. Now, we declare that we gauge the Weyl symmetry and discard $\varphi$. To make the action simpler, we choose $\xi = -2c_2$ so that the Palatini scalar curvature $\mathcal{R}$ does not couple to $|\Phi|^2$. This choice is particularly useful because then the variation of the Palatini connection $\Gamma^{\mu}_{\rho\sigma}$ just gives
the condition that the Palatini connection is identified with the Christoffel connection. With this choice, $\mathcal{R} = R$ as a consequence of the equations of motion and we recover the Einstein action with cosmological constant coupled with the Higgs field $|\Phi|^2$.\footnote{When the Palatini curvature has non-minimal couplings to the matter, the Palatini connection is not necessarily equal to the Chirstoffel connection \cite{Iglesias:2007nv}.}

This formulation of the standard model may be of esthetic interest because the starting point of the action is completely dimension free. In addition, our formulation has a prediction that the non-minimal gravitational coupling of the Higgs field is fixed to be $1/6$. 

In order to obtain the full standard model of the particle physics within our setup, we need to introduce gauge fields and fermion fields.
The introduction of fermion fields requires a little bit of care because (1) it uses a spin connection and (2) we probably cannot introduce the mass for the fermions. With the fermion fields $\Psi$ and the gauge fields, the action of the matter is schematically given by  
\begin{align}
\int d^4x \sqrt{-g}\left( \frac{1}{2g^2_{\mathrm{YM}}}\,\mathrm{Tr} F_{\mu\nu} F^{\mu\nu}  + \theta \,\mathrm{Tr} \epsilon^{\mu\nu\rho\sigma}F_{\mu\nu} F_{\rho\sigma}  +  \bar{\Psi} D_\mu \gamma^\mu \Psi + y(\Psi \Psi \Phi) + \bar{y}(\bar{\Psi}\bar{\Psi}\bar{\Phi}) + \text{Higgs} \right) . \label{other}
\end{align}
The Higgs sector is essentially given in \eqref{Higgs} by replacing the derivative $\partial_\mu$ with the gauge covariant one. Without violating the Weyl symmetry, one may use the standard spin connection constructed out of the vielbein  $e_{\mu}^a$ for fermion kinetic terms.
More precisely, we introduce the torsion-free Levi-Civita spin connection constructed out of the vielbein:
\begin{align}
\omega_{\mu}^{ab} = \frac{1}{2}e^{\nu a}(\partial_\mu e_\nu^b - \partial_\nu e_{\mu}^b) -\frac{1}{2}e^{\nu b} (\partial_\mu e_\nu^a - \partial_\nu e_{\mu}^a) -\frac{1}{2} e^{\rho a} e^{\sigma b}(\partial_\rho e_{\sigma c} - \partial_\sigma e_{\rho c}) e_{\mu}^c \ .
\end{align}
We also introduce the Palatini spin connection $\Omega_{\mu}^{ab}$ and we treat it as an independent variable.

Under the Weyl transformation, the vielbein transforms as $e_{\mu}^a \to \Omega(x) e_{\mu}^a$ so that $g_{\mu\nu} = e_{\mu}^a e_{\nu a} \to \Omega^2(x) g_{\mu\nu}$. Then the Levi-Civita spin connection transforms as
\begin{align}
\omega_{\mu}^{ab} \to \omega_{\mu}^{ab} - e^{\nu a} e_{\mu}^b \partial_\nu (\ln\Omega) + e^{\nu b} e_{\mu}^a \partial_\nu (\ln \Omega) 
\end{align}
while the Palatini spin connection transforms as
\begin{align}
\Omega_{\mu}^{ab} \to \Omega_{\mu}^{ab} \ .
\end{align}

In the gravitational part of the action, the Weyl invariant curvature term i.e. $\mathcal{R}^2$ is now constructed out of the Palatini spin connection $\Omega_{\mu}^{ab}$. 
\begin{align}
\mathcal{R} = e^\mu_a e^\nu_b (\partial_\mu \Omega_{\nu}^{ab} -\partial_\nu \Omega_{\mu}^{ab} + \Omega_{\mu}^{ac} \Omega_{\nu c}^{\ \  b} - \Omega_{\nu}^{ac} \Omega_{\mu c}^{\ \  b})
\end{align}
The non-minimal coupling to the Higgs field contains both the Palatini curvature and the Ricci scalar as in \eqref{Higgs}. 
On the other hand, in the fermionic kinetic term, we exclusively use the Levi-Civita spin connection $\omega^{ab}_{\mu}$ (in addition to the Yang-Mills gauge field $A_\mu$) as
\begin{align}
D_\mu \Psi = (\partial_\mu + iA_\mu) \Psi + \frac{1}{2}\omega_{\mu}^{ab}\Sigma_{ab} \Psi \ 
\label{covariant}
\end{align}
where $\Sigma_{ab}=\frac{1}{2}[\gamma_a,\gamma_b]$. In \reff{covariant} we use $\omega^{ab}_{\mu}$ rather than the Palatini spin connection $\Omega_{\mu}^{ab}$ in order to preserve the Weyl invariance because we assume the Palatini spin connection does not transform under the Weyl transformation (like the Palatini connection $\Gamma^{\mu}_{\rho\sigma}$) so the Weyl transformation of the fermion must be cancelled by the Levi-Civita spin connection. 

Phenomenologically this is very appealing because if we used the Palatini spin connection in the fermionic kinetic terms (as we usually do in the conventional Palatini formulation), the equations of motion for the Palatini spin connection $\Omega^{ab}_{\mu}$  would dictate that it has an additional torsional contribution from the fermionic spin currents, resulting in a spin-spin interaction in the final matter action as in the Einstein-Cartan theory. In our case, after rewriting the action in the equivalent Einstein form, the variation of the Palatini spin connection makes it identified with the Levi-Civita spin connection and hence
we do not generate such (so-far) unobserved four-fermi interactions. 

Note also that the Weyl invariance does not allow fermion mass terms or cubic self-interaction terms for scalar fields (even if they were allowed by gauge symmetry in the more general matter action). In this sense, it is remarkable that our standard model of particle physics as well as all the gravitational physics we observe in the universe is compatible with our formulation of Weyl invariant Palatini gravity.
\section{Conclusion}
In this work, we showed that the Palatini formulation of pure $R^2$ gravity is equivalent to Einstein gravity with non-zero cosmological constant. No massless scalar field appears in contrast to the metric formalism where it appears as a Nambu-Goldstone boson \cite{YN3}. We gauged the Weyl symmetry and the gauge symmetry is spontaneously broken. There is no Nambu-Goldstone boson for spontaneously broken gauge symmetry. This is a positive development because there is no evidence of such a massless scalar in nature (including cold dark matter which is expected to be massive). The equivalence is of aesthetic appeal because the original action is Weyl invariant and has no scale.  

We then construct a Weyl invariant unitary action where we include a non-minimally coupled massless Higgs field. The Weyl invariance requires a hybrid action that includes both the Palatini curvature scalar $\mathcal{R}$ and the usual Ricci scalar $R$.   We show that this action is equivalent to Einstein gravity with cosmological constant and a non-minimally coupled massive Higgs with coefficient fixed to be $1/6$.  So the theory, besides its aesthetic appeal, makes a non-trivial prediction that the coupling of the Higgs to the Ricci scalar comes with a coupling constant of $1/6$. In other words, the original Weyl invariance which is subsequently spontaneously broken, leaves an imprint on the final action and determines one of the coupling constants. This is pertinent for work on inflationary models which couple the Higgs boson to curvature in Einstein gravity (as a recent example see \cite{Calmet}). Typically, in such work, one considers the coupling to be an arbitrary constant $\xi$ instead of $1/6$. 

Remarkably, one can incorporate into this paradigm all the other standard model fields, including fermions. We showed that this can be accomplished by working with two spin connections instead of only one: the torsion-free Levi-Civita connection $\omega_{\mu}^{ab}$ constructed out of the vielbein and the Palatini spin connection $\Omega_{\mu}^{ab}$ as an independent variable. The former is used for the fermions and the latter for the $\mathcal{R}^2$ gravitational action. Therefore our final action reproduces the standard model as we know it (with the added benefit of fixing the coupling constant between the Higgs and the Ricci scalar to be $1/6$).
 
\section*{Acknowledgments}
Y. N. is in part supported by JSPS KAKENHI Grant Number 17K14301. A.E. is supported by an NSERC discovery grant.

\end{document}